\documentclass[pre,aps,twocolumn,showpacs,floatfix,superscriptaddress,nobibnotes,letterpaper,amsfonts,amssymb,amsmath]{revtex4}
\usepackage{graphicx}
\usepackage{pdfpages}
\usepackage{dcolumn}
\usepackage{bm}
\usepackage{color}
\usepackage{sidecap}
\usepackage{float}
\usepackage{hyperref}
\begin{document}

\title{Knot formation of dsDNA pushed inside a nanochannel}

\author{Jan Roth\"{o}rl}
\affiliation{Institut f\"{u}r Physik, Johannes Gutenberg-Universit\"{a}t, Staudinger Weg 9, D-55099 Mainz, Germany}
\author{Sarah Wettermann}
\affiliation{Institut f\"{u}r Physik, Johannes Gutenberg-Universit\"{a}t, Staudinger Weg 9, D-55099 Mainz, Germany}
\author{Peter Virnau}
\thanks{Authors to whom the correspondence should be addressed: Aniket.Bhattacharya@ucf.edu, virnau@uni-mainz.de}
\affiliation{Institut f\"{u}r Physik, Johannes Gutenberg-Universit\"{a}t, Staudinger Weg 9, D-55099 Mainz, Germany}
\author{Aniket Bhattacharya}
\thanks{Authors to whom the correspondence should be addressed: Aniket.Bhattacharya@ucf.edu, virnau@uni-mainz.de}
\affiliation{Department of Physics, University of Central Florida, Orlando, Florida 32816-2385, USA}

\begin{abstract}
Recent experiments demonstrated that knots in single DNA strands can be formed by hydrodynamic compression in a nanochannel. In this letter, we further elucidate the underlying molecular mechanisms by carrying out a compression experiment {\em in silico}, where an equilibrated coarse-grained double-stranded DNA confined in a square channel is pushed by a piston. 
The probability of forming knots is a non-monotonic function of the persistence length and can be enhanced significantly by increasing the piston speed. 
Under compression, knots are abundant and delocalized due to a backfolding mechanism from which chain-spanning loops emerge, while knots are less frequent and only weakly localized in equilibrium. Our {\em in silico} study thus provides insights into the formation, origin and control of DNA knots in nanopores.
\end{abstract}


\maketitle

In many biological processes, a double-stranded DNA (dsDNA) is confined in a geometry much shorter than its contour length in a highly organized and compact state and often under high pressure~\cite{Miceletti-Review,Reisner-Review}. A classic example is an organized state of a dsDNA strand in a viral capsid~\cite{ArsuagaPNAS2002,ArsuagaPNAS2005,Petrov,MarenduzzoPNAS2009,Reith:NAR:2012,Marenduzzo:PNAS:2013}. The viral DNA uses the stored elastic energy for its invasion process. Intriguingly, this DNA was found to be highly knotted particularly in a mutant variant for which both sticky ends are allowed to reside within the capsid~\cite{ArsuagaPNAS2002,ArsuagaPNAS2005}.
It is in general difficult to develop an experimental protocol to study an actual system {\em in vitro}, although there have been studies to measure the force and the organized topology of the dsDNA inside a capsid~\cite{SmithPNAS2014,SmithNatPhys2016}. During the last decade advancements in nanotechnology have enabled us to prepare nanochannels of sub-persistence length dimensions~\cite{Saltextension}. DNA pushed inside nanofluidic devices~\cite{Stein,Nanogroove,AhmedPRL,AhmedMacro,Reisner-NatureCommn-2018} is now used for mapping genomes, sequence motifs, structural variations~\cite{Hancao,Ramseyfunnel}.\par
Recently, nanochannels were even used for the detection of knots in DNA~\cite{Plesa_NatureNano_2019}. Due to a controllable and simpler geometry, nanochannels offer immense promise to understand universal aspects of biological phenomena using well established concepts from polymer physics~\cite{Sakaue-cavity,Jun2008-compression}. Besides problems of biological significance and of human health, nanochannel based experiments claimed the occurrence of jamming ~\cite{SmithPNAS2014,SmithNatPhys2016}- which indicates that confined bio-polymers offer yet another platform to study slow relaxation and glassy dynamics. Thus studies of chain compression in nanochannels appeal to broad areas of science.\par
Many numerical studies of knots have established numerous results on generic~\cite{Micheletti_PhRep2011,Vologodskii:1974,Koniaris:prl:1991,Mansfield:Macro:1994,Grosberg:PRL:2000,Marcone_2005,Virnau_JACS_2005,Foteinopoulou:PRL:2008,Orlandini_JBP_2013,Miceletti-ACSMacro-2014,Trefz_PNAS_2014,Miceletti-SoftMatter2017,Meyer:ACSMacro:2018:knots_melt,Zhang_2020,Tubiana_2021} and biopolymers~\cite{Miceletti-Review,MarenduzzoPNAS2009,Marenduzzo:PNAS:2013,Rieger:PLoS:2016,Virnau_PLoScb_2006,Sulkowska_PNAS_2008,Sulkowska_PNAS_2009,Boelinger_PLoS_2010,knotprot,Wuest:PRL:2015}.
While knots are, e.g., abundant in single ideal chains~\cite{Vologodskii:1974,Koniaris:prl:1991}, the addition of excluded volume typically reduces the knotting probability significantly~\cite{Koniaris:prl:1991,Virnau_JACS_2005}, while spherical confinement or globular states enhance knotting~\cite{Mansfield:Macro:1994,Virnau_JACS_2005}. For the latter, knots tend to be delocalized, i.e. average sizes scale linearly with the chain length~\cite{Virnau_JACS_2005}, while knots are weakly localized and scale sub-linearly for under ideal or good solvent conditions~\cite{Marcone_2005,Virnau_JACS_2005}.\par
Previously, we have studied compression of semi-flexible polymers in nanochannels using a Langevin dynamics (LD) scheme~\cite{Huang-Polymer2016,Bernier-MM2018}. These LD simulation studies and another recent study~\cite{SakaueOdijkCompression} have provided substantial insights about many details at smaller length scales unattainable experimentally, but are essential for microscopic understanding and interpretation of nanochannel experiments using fundamental laws of physics. One of the advantages of these simulation studies is that, unlike an actual experiment, one can vary confining dimensions and chain stiffness easily and is thus capable of extending the simulation studies for a broader parameter space which is often quite expensive to design experimentally. In our recent LD simulation study~\cite{Bernier-MM2018} we mimicked a recent experiment {\em in silico} where dsDNA - modeled as a semi-flexible polymer - was pushed inside a rectangular open-ended nanochannel much longer than required to attain a steady state. By varying the bending rigidity of the chain we showed how the structure evolves from a disordered state to a highly organized spooled state. Furthermore, the LD simulation revealed a detailed picture of how the fold nucleation originates at the piston end and expands during the compression process~\cite{Bernier-MM2018}.\par
An important and relevant question in these compression studies in the biological context is to study how the formation of knots are initiated and once formed how do they spatially evolve under confinement. Theoretical and simulation studies have been further fueled by a recent experiment that demonstrates that knots indeed occur in compression experiments~\cite{Reisner-NatureCommn-2018}. Micheletti and coworkers have studied knot formation of cyclical DNA in confined geometry and demonstrated the occurrence of knots from LD simulation studies. They further showed that the knotting probability is non-monotonic as a function of the bending rigidity~\cite{Orlandini_JBP_2013,Miceletti-ACSMacro-2014,Miceletti-SoftMatter2017}. In this letter, we extend our analyses of LD studies to investigate knot formation when a confined dsDNA is being pushed by a nano-dozer in a quasi-one-dimensional nanochannel whose width is much smaller than the contour length of the dsDNA.\par
\begin{figure}[ht!]
\includegraphics[width = 8.4cm]{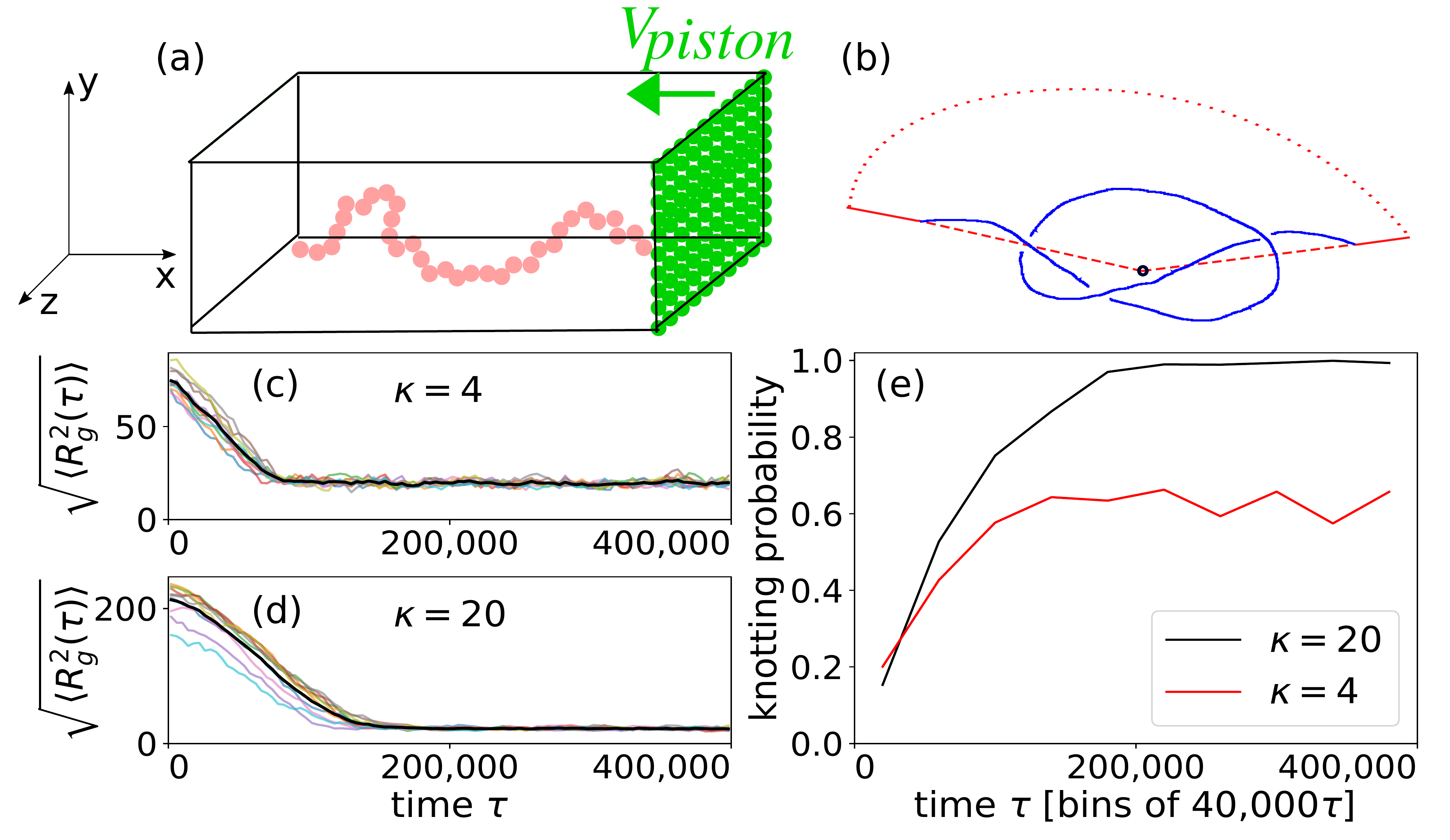}
\caption{Schematics of the simulation model. (a) A semi-flexible bead-spring chain confined inside a rectangular nanochannel is pushed by the green piston from right to left at velocity  $v=v_{\mbox{\scriptsize piston}}$. The confinement potentials are imposed on four (two $xy$ and two $xz$) planes, and by the moving piston in the $yz$ plane in the negative $x$ direction. The chain is free to move on the side opposite the piston. (b) Demonstration of the knot closure. The termini of the polymer are connected with the center of mass (black dot) as indicated by the dashed red lines. The solid red lines are then appended to the polymer and connected by a closing arc drawn with red dots. (c,  d) The root-mean-squared radius of gyration  $\sqrt{\langle R_g^2(\tau) \rangle}$ of the polymer as a function of LD time $\tau$ for 10 simulations each and two different values of the chain stiffness, $\kappa = 4.0$ and 20, respectively. The polymer is compressed until it reaches a steady state with a constant radius of gyration. At this point it moves like a blob of a fixed shape. The steady state $\sqrt{\langle R_g^2  \rangle}$ for the polymer with lower stiffness is significantly smaller in its final state. (e) The knotting probability as a function of time $\tau$. During compression, the knotting probability increases while it is constant in the steady state. A higher stiffness leads to a higher knotting probability in the steady state. The results are averaged over 20 independent runs each.}
\label{Model}
\end{figure}
An important difference from the previous studies is that our system is open ended in one direction and that we study the evolution of knots until the system approaches the steady state.
The key result is that the confined chain in the nanaochannel pushed by a nano-dozer will progressively become highly knotted with delocalized knots. The knotting probability is greatly enhanced compared to corresponding equilibrium simulations, which in addition to compactification can be traced back to a backfolding mechanism for semi-flexible chains. 
Next, we describe the model, some essential facts about the LD simulation scheme, how our coarse-grained chains can be mapped onto DNA and the method that we use to analyze knots.\par
{\bf Coarse-grained polymer model:}~The CG model of of polymer for BD simulation used here is exactly the same as in our previous
publication~\cite{Bernier-MM2018} where a bead-spring model polymer chain
is confined to an open-ended rectangular channel and pushed from the right with a piston in the negative $x$-direction (Fig.~\ref{Model}(a)). 
The semi-flexible chain (Fig.~\ref{Model}(b)) is represented by a generalized bead-spring model~\cite{Grest}. 
The details are in the supplementary section. 
The chain persistence length is controlled by varying the stiffness parameter in the bond-bending potential $\ell_p = \frac{\kappa \sigma}{k_B T}$ 
where $k_B$ is the Boltzmann constant and $T$ is the temperature. \par 
{\bf The Langevin dynamics simulation:}~The numerical integration is implemented using the algorithm introduced by Gunsteren and Berendsen~\cite{Gunsteren}. Our previous experience with LD simulations suggest that appropriate parameter specifications are $\gamma = 0.7\sqrt{m\epsilon/\sigma^2}$,
$k_\textrm{FENE} = 30\epsilon/\sigma$, $R_0 = 1.5\sigma$, and a temperature $k_B T/\epsilon = 1.2$. For a time step $\Delta t = 0.01$ these parameter values produce stable trajectories over a very long period of time and do not lead to an unphysical crossing of a bond by a monomer~\cite{Huang_JCP_2014,Huang_JCP_2015}. The average bond length stabilizes at $b_l = 0.970 \pm 0.002 $ with negligible fluctuation regardless of chain size and rigidity~\cite{Huang_JCP_2014}.
The piston is moved with a constant velocity of $v_0 = 0.005$ if not noted otherwise after an initial equilibration of the chain.
We ensure that the MD time for the pushing phase is long enough for the chain to attain a steady state shown in Fig.~\ref{Model}(c)-(d) that displays the connection between chain extension along the channel axis (Fig.~\ref{Model}c) and knot formation (Fig.~\ref{Model}d) in approach to the steady state. Times to reach the latter depend on bond stiffness $\kappa$ as seen from the behavior of $\langle \sqrt{\langle R_g^2(t)} \rangle $ in Fig.~\ref{Model}c-d. While for $\kappa = 4$ reaching a steady state takes less than $50,000\tau$, it takes around $160,000\tau$ for $\kappa = 20$. 
For each $\kappa$ and $v$, physical quantities are averaged over at least ten independent runs.\par
{\bf Reptation Monte Carlo simulation to study the equilibrium limit:}~
Note that piston speeds in coarse-grained implicit solvent simulations are typically orders of magnitude faster when compared to experiments. Therefore, we have also undertaken reptation Monte Carlo simulations of a slightly simplified model system with fixed bond lengths and hard walls. These simulations allow for a comparison of our dynamical investigations with equilibrium values (corresponding to piston velocity $v\rightarrow0$).\par
\begin{figure*}[ht!]
\includegraphics[width = 17.2cm]{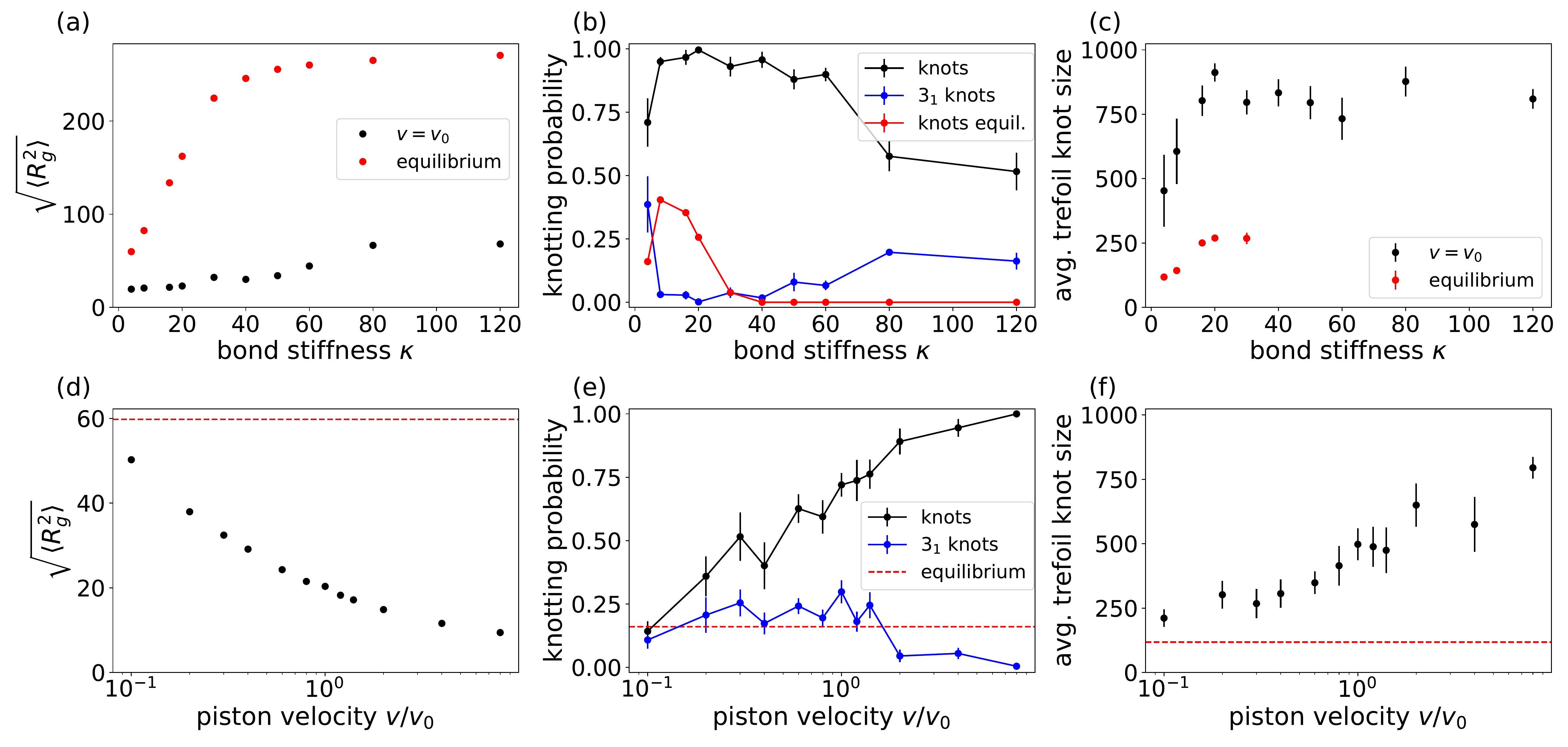}
\caption{(a) Root-mean-squared radius of gyration $\sqrt{\langle R_g^2(t) \rangle}$ for chains of different stiffness (b) Knotting probability and occurrence probability of trefoil knots for different values of the bond stiffness $\kappa$. Error-bars are determined by taking the standard deviation over the square-root of the number of runs. Lines are added for readability. Knotting probability for $v = 0$ is strictly zero for $\kappa \geq 50$. (c) Trefoil knot lengths found for different $\kappa$ with a moving piston and in equilibrium. (d) $\sqrt{\langle R_g^2(t) \rangle}$ for  different piston velocities at $\kappa = 4$. The red dashes indicate the average result in MC simulations without a moving piston. (e) Knotting probabilities for different piston velocities. (f) Trefoil knot lengths for different piston velocities.}
\label{fig:Knots}
\end{figure*}\par 
{\bf Knot Analysis:}~
Knots in a closed chain are typically characterized by the minimum number of crossings observed when projecting a 3D chain onto a plane and can be considered as a fine gauge for the overall structure. Apart from the unknotted ring, the so-called unknot, the simplest knot is the trefoil $(3_1)$ knot, which contains three crossings. There is one knot type with four crossings ($4_1$) and two with five crossings, and from there on the number of different knots with the same number of crossings increases exponentially. In our setup the polymer chain is open, and therefore, a closure connecting both ends of the chain has to be defined. First, we connect the end-points of each polymer with its center of mass. Along these lines we define a closure which emerges from one terminus follows the first line connects to the second one far away from the polymer and ends at the second terminus~\cite{Virnau_PLoScb_2006}. 
After closure, the Alexander polynomial can be determined as described in detail in~\cite{Virnau_Phys_Proc_2010} (compare Fig.~\ref{Model}b). Knot sizes are determined by successively removing monomers from the ends of a polymer until the knot type changes ~\cite{Virnau_JACS_2005}.\par
{\bf Mapping onto DNA and comparison with experiments:}~
Mapping our semi-flexible chain onto DNA is based on the equation $\ell_p=\kappa\sigma/k_BT$. For simplicity, we assume a solvent-independent persistence length of 50~nm or 150 base pairs. Furthermore, we assume that our beads describe the locus of a double-stranded DNA strand. In high salt conditions (1M NaCl), charges of DNA are completely screened and $\sigma\approx2.5~nm$. In physiological conditions charges are only partially screened and $\sigma\approx5~nm$, and for low salt conditions $\sigma$ increases even further to about $15nm$ at 0.01 M NaCl~\cite{Rybenkov_1993,Trefz_PNAS_2014,Rieger:PLoS:2016}. With a simulation temperature of $T=1.2$ used throughout we obtain (in simulation units) $\kappa=24$ for high salt, $\kappa=12$ for physiological and $\kappa=4$ for low salt conditions. 
This allows us to put our simulations in the context of recent experiments by Amin et al~\cite{Reisner-NatureCommn-2018} undertaken at an estimated ionic strength of $8~mM$ which corresponds to our low salt scenario. Our chain has a contour length of $L=N\sigma = 1024\sigma=15,360~nm$ or 46,080 base pairs, while our confining tube has a width of $16\sigma \approx 240~nm$. This compares to 168,903 base pairs and tube dimensions of $325 \times 415~nm$ used in Ref.~\cite{Reisner-NatureCommn-2018}. Note that the mapping changes drastically with ionic conditions.\par
\begin{figure}[ht!]
\includegraphics[width = 8.6cm]{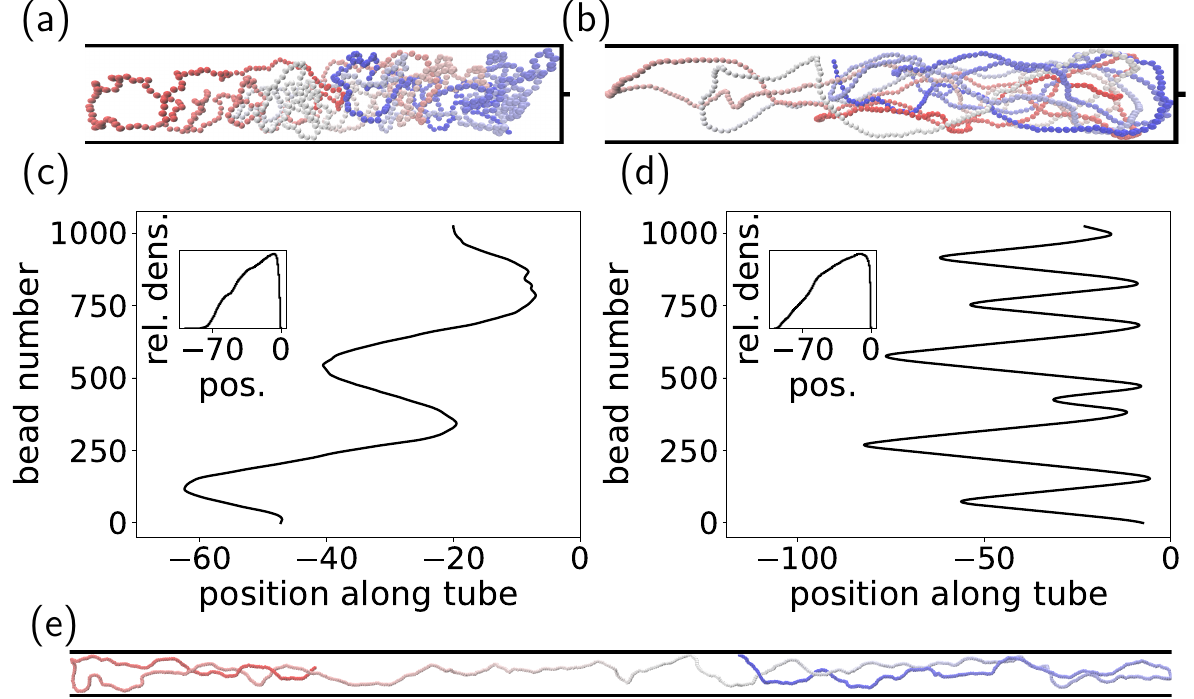}
\caption{(a, b) Structures for $\kappa = 4$ (a) and $\kappa = 20$ (b) visualized using VMD~\cite{VMD}. Beads are colored from blue to red according to their monomer number. (c, d): The plots show the average position along the tube for all beads averaged over a simulation time of $50,000 \tau$ for simulations at $\kappa = 4$ (c) and $\kappa = 20$ (d) with the piston at position 0. The insets show the relative density of beads along the structure averaged over the same frames. The structure for $\kappa = 20$ is significantly larger. The highest density for both structures is found to be close to the piston on the right. (e) Equilibrium structure without piston for $\kappa = 20$.}
\label{fig:Contacts}
\end{figure}

{\bf Results:}~
Fig.~\ref{fig:Knots} summarizes the main findings of our study. Applying a pushing force leads to a compactification of the polymers (Fig.~\ref{fig:Knots}a), which in turn dramatically increases the occurrence of knots in the steady state in comparison to equilibrium values (Fig.~\ref{fig:Knots}b).
Likewise, the amount of trefoil knots is reduced for configurations with a higher total knotting probability because the high density induces the formation of multiple or more complex knots.
In this compact state, knots are delocalized and span over the whole chain as indicated for the example of trefoil knots in Fig.~\ref{fig:Knots}c where for $\kappa > 20$ the average length of the knot is approximately 80\% of the contour length or higher, which implicates that knots are formed preferentially near each end (please see Fig.~\ref{fig:Contacts}e), while knots in equilibrium conformations are significantly smaller. These findings in a sense mirror previous observations, e.g. in Ref.~\cite{Virnau_JACS_2005}, which demonstrated that a $\theta$-transition from a swollen coil to a globular state is not only accompanied by an increase in knotting but also by a delocalization of the latter. Fig.~\ref{fig:Knots}d investigates the influence of the piston velocity for the experimentally relevant case of $\kappa=4$. Again, compactification with increasing velocity is directly related to an increase of overall knotting. These results suggest that the occurrences of knots can be tuned by the speed of the piston and converge towards the equilibrium values for small piston velocities (Fig.~\ref{fig:Knots}e). As indicated above the decrease of knotting towards the equilibrium state at slow piston velocities is again accompanied by a trend towards a weak localization of trefoil knots (Fig.~\ref{fig:Knots}f)~\cite{Virnau_JACS_2005}.\par
Fig.~\ref{fig:Contacts} sheds light on these findings from a molecular basis. For $\kappa=4$ the structure is disordered but the position of the monomers is still correlated with their sequence as indicated by the color scheme in  Fig.~\ref{fig:Contacts}a and the bead positions in Fig.~\ref{fig:Contacts}c. For $\kappa = 20$, the persistence length already exceeds the width of the tube which in conjunction with compactification leads to backfolding (Figs.~\ref{fig:Contacts}b, d). The backfolding on the other hand creates loops which are a prerequisite for knots and in turn explains the initial rise in knotting with $\kappa$ as well as their delocalization.
For large persistence lengths, backfolding becomes more difficult resulting in a lower knotting probability (Fig.~\ref{fig:Knots}b).
In the equilibrium case, the compactification from the piston is no longer present and for $\kappa=20$, the chain can already spread throughout the channel which leads to a low knotting probability and weakly localized knots (Fig.\ref{fig:Contacts}e).\par
{\bf Conclusions:~} 
In this letter we investigate velocity induced knot ``production'' in a nanochannel in comparison to those under equilibrium conformations. Both knotting probability and knot sizes depend strongly on piston velocity and resulting compactification as well as chain stiffness which can be, e.g., mitigated by adjusting ionic conditions and screening of charges in DNA. We observe that for chain stiffness greater than the width of the nanochannel knots form by the backfolding mechanism. Since backfolding becomes harder for larger stiffnes, the probability of knot formation decreases which explains the observed nonmonotonic characteristic of knot formatation in a nanochannel. We also study relative occurrences of  complex knots as a function of the piston velocity and the chain stiffness. Our study thus sheds some new light on recent experiments in which DNA knots were created in a flow channel~\cite{Reisner-NatureCommn-2018} and provides insight on the molecular origin and control of self-entanglements under these conditions.
\par
\begin{acknowledgements}
AB thanks Aiqun Huang for the preliminary runs. The authors acknowledge partial funding from TopDyn. We are grateful to the Deutsche Forschungsgemeinschaft (DFG, German Research Foundation) for funding this research: Project number 233630050-TRR 146. The LD simulation were carried out at the UCF's high performance computing cluster STOKES. The authors gratefully acknowledge the computing time granted on the supercomputer Mogon offered by the Johannes Gutenberg University Mainz (hpc.uni-mainz.de), which is a member of the AHRP (Alliance for High Performance Computing in Rhineland Palatinate,  www.ahrp.info) and the Gauss Alliance e.V. AB acknowledges travel support from 
Institut f\"{u}r Physik, Mainz.
\end{acknowledgements}
\vfill
\bibliographystyle{apsrev4-1}
\bibliography{references}
\end{document}